\documentclass[12pt,journal, onecolumn, draftcls]{IEEEtran}
\usepackage{amsmath}
\usepackage{graphicx}
\usepackage{epsfig}
\usepackage{psfrag}
\usepackage{array}
\usepackage{amssymb}

\usepackage{amsmath,graphicx}
\usepackage{epsfig}
\usepackage{amssymb}
\usepackage{hyperref}
\usepackage{color}
\usepackage{subfigure}
\usepackage{amsthm}

\begin{document}

\title{How Many Beamforming Vectors Generate the Same Beampattern?}
\author{Arash~Khabbazibasmenj, Aboulnasr Hassanien, and Sergiy~A.~Vorobyov

\thanks{The authors are with the Dept. of Electrical and Computer 
Engineering, University of Alberta, Edmonton, Alberta, T6G~2V4, Canada 
(emails: khabbazi@ualberta.ca; hassanie@ualberta.ca; svorobyo@ualberta.ca). 
S.~A.~Vorobyov is on leave and currently with the Dept. Signal Processing 
and Acoustics, Aalto University, Finland.

This work is supported in part by the Natural Science and Engineering Research
Council (NSERC) of Canada.

} } 

\maketitle

\begin{abstract}
In this letter, we address the fundamental question of how many
beamforming vectors exist which generate the same beampattern?
The question is relevant to many fields such as, for example,
array processing, radar, wireless communications, data compression,
dimensionality reduction, and biomedical engineering. The
desired property of having the same beampattern for different
columns of a beamspace transformation matrix (beamforming vectors)
often plays a key importance in practical applications. The result
is that at most $2^{M-1}-1$ beamforming vectors with the same
beampattern can be generated from any given beamforming vector.
Here $M$ is the dimension of the beamforming vector. At
the constructive side, the answer to this question allows for
computationally efficient techniques for the beamspace transformation
design. Indeed, one can start with a single beamforming vector,
which gives a desired beampattern, and generate a number of
other beamforming vectors, which give absolutely the same
beampattern, in a computationally efficient way. We call the initial
beamforming vector as the mother beamforming vector. One possible
procedure for generating all possible new beamforming vectors with
the same beampattern from  the mother beamforming vector is
proposed. The application of the proposed analysis to the
transmit beamspace design in multiple-input multiple-output
radar is also given.
\end{abstract}


\section{Introduction}
Beamspace transformation \cite{Anderson}, \cite{Foster} and
beamforming \cite{Vorobyov} techniques are the key approaches,
among others, in array signal processing \cite{VanTrees}-\cite{Zoltowski},
radar \cite{Farina}, multiple-input multiple-output (MIMO) radar
\cite{Fuhrman}-\cite{Khabbazi}, wireless communications
\cite{Tse}-\cite{Molish}, data compression and dimensionality reduction
\cite{Jieping}-\cite{DimRed}, biomedical engineering \cite{Rivera}, etc.

In the traditional applications in array processing and dimensionality
reduction, it is often desirable to reduce the high dimensional space
into a lower one by means of the beamspace transformations. In
more recent applications to MIMO radar, it has been required not
only to design a lower dimensional transmit beamspace but also to
transmit a number of orthogonal waveforms from a larger number
of transmit antenna elements while achieving transmit coherent
processing gain. While designing such a transmit beamspace,
certain properties have to be satisfied such as a uniform power
distribution for different transmit waveforms in the desired
sector where the targets are likely to be located \cite{Hassanien4}. 
The latter enables, for example, to enforce at the transmitter 
the very useful rotational invariance property \cite{Roy}, 
\cite{KHVM} which can significantly simplify and improve, for 
example, the direction-of-arrival (DOA) estimation techniques used 
at the receive antenna array.

From a practical point of view, generating a transmit beamspace that
satisfies a number of properties is very desirable. Thus, it is of 
interest to look for simple design techniques allowing to start with 
a single beamforming vector, which we call the mother beamforming 
vector by analogy with mother wavelet \cite{WaveletBook} in wavelet 
analysis, and generate a number of other beamforming vectors that all 
have the same beampattern as the mother beamforming vector. Beamforming
vectors that possibly satisfy some additional practically important
properties can then be selected from a set of so generated
beamforming vectors. In wavelets, self-similarity is an important
property where basis functions are all obtained from a single
prototype mother wavelet using scaling and translation. It is
interesting that a similar property exists also in the beamspace
design problem. To the best of the authors knowledge, this
property has not been known or exploited before.

The contributions of this letter are as following.
\begin{itemize}
\item We first address the fundamental question of how
many beamforming vectors which generate the same beampattern
as the mother beamforming vector exist.
\item At the constructive side, we develop a computationally
efficient technique for generating all such beamforming vectors.
\item The application of the proposed analysis to the transmit
beamspace design in MIMO radar is also given.
\end{itemize}

The rest of the letter is organized as follows. The main results on
the number of beamforming vectors, which have the same
beampattern as the mother beamforming vector, and on the procedure
of constructing such beamforming vectors are given in Section II.
Section III is devoted to the application of the proposed analysis
to the transmit beamspace design in MIMO radar, which aims at
demonstrating the practical usefulness of the results obtained in
Section II. The letter is completed with conclusion.

\section{Main Results}
Let us consider a uniform linear array (ULA) with $M$ antenna elements.
The steering vector of the array towards direction $\theta$ is denoted
as $\mathbf a (\theta)$. The transmit array beampattern can be expressed
as
\begin{equation}
p(\theta) =\| \mathbf w^H {\mathbf d}(\theta) \|^2
\label{beampattern}
\end{equation}
where $\mathbf w$ is a beamforming vector, ${\mathbf d}(\theta) =
{\mathbf a}^{\ast}(\theta)$, and $\| \cdot \|$ and $(\cdot)^{\ast}$ 
stand for the Euclidian norm of a vector and conjugation, respectively. 
Let the beampattern corresponding to a given beamforming vector 
$\mathbf w$, referred to as the mother beamforming vector,
satisfies certain shape design requirements, but it does not satisfy
other practically important requirements. Such a requirement is, for
example, a uniform power distribution across the antenna elements.
The question then arises about existence of other distinct
beamforming vectors which generate the same exact beampattern as the
mother beamforming vector $\mathbf w$ and which in addition satisfy
other possible design requirements. The following theorem states the
total number of distinct beamforming vectors with the same exact
beampattern as the mother beamforming vector $\mathbf w$.

\textbf{Theorem~1 :}
For an arbitrary transmit beamforming vector $\mathbf w$ that
corresponds to a ULA of size $M$, there are at most $2^{M-1}-1$
other different beamforming vectors which have the same exact
beampattern as $\mathbf w$. A constructive solution for obtaining
all possible beamforming vectors is given in the proof.

{\it Proof :} Let $\mathbf w$ and $\mathbf v$ be two beamforming
vectors with the exact same transmit beampattern. The latter implies
that
\begin{equation}
\mathbf w^H \mathbf D(\theta) \mathbf w = \mathbf v^H \mathbf
D(\theta) \mathbf v, \quad \quad  \forall \theta \in [-\pi/2,\pi/2]
\label{eq1}
\end{equation}
where the matrix $\mathbf D(\theta)$ is defined as $\mathbf D
(\theta) \triangleq \mathbf d(\theta) \mathbf d^H(\theta)$. It can
be easily verified that the matrix $\mathbf D(\theta)$ has Toeplitz
structure.

In the rest of this proof, we will refer to the descending
diagonals of a matrix from left to right as the first, second, and up
to $(2M+1)$th off-diagonal, respectively. For the notation simplicity,
let us define the following vector
\begin{eqnarray}
\mathbf z(\theta) \!\!\!\!\!&\triangleq&\!\!\!\!\! \bigg( e^{j 2 \pi 
(M-1) d \sin (\theta) }, \cdots, e^{j 2 \pi (2) d \sin (\theta)}, e^{j
2 \pi d \sin (\theta)}, 1, e^{-j 2 \pi  d \sin (\theta)},\cdots, e^{-j 
2 \pi d (M-1) \sin (\theta)} \bigg)^T
\end{eqnarray}
whose $i$th element is equal to $i$th off-diagonal elements of $\mathbf
D (\theta)$. Moreover, we will express the Toeplitz matrix $\mathbf D
(\theta)$ with the diagonal and off-diagonal elements defined in $\mathbf
z(\theta)$ as $\mathbf T(\mathbf z(\theta))$ where $\mathbf T(\cdot)$ is 
an operator which generates a Toeplitz matrix.

Since equation \eqref{eq1} holds valid in all directions, the following 
set of equations can be resulted by fixing $\theta$ to an arbitrary set 
of angles dented as $\theta_{k}, \ k=1,\cdots,2M+1$
 \begin{equation}
\mathbf w^H \mathbf D(\theta_{k}) \mathbf w = \mathbf v^H \mathbf
D(\theta_{k}) \mathbf v, \quad \quad k=1,\cdots,2M+1 .
\label{AuxEq}
\end{equation}
By linearly combining the set of equations in \eqref{AuxEq} using an 
arbitrary set of coefficients denoted as $c_k, \ k=1,\cdots,2M+1$, 
the following equality can be concluded
\begin{equation}
\mathbf w^H \bigg(\sum_{k=1}^{2M+1} c_k \mathbf D (\theta_{k}) \bigg)
\mathbf w = \mathbf v^H \bigg(\sum_{k=1}^{2M+1} c_k \mathbf D
(\theta_{k})\bigg) \mathbf v \label{LinComb}
\end{equation}

In what follows,
we will show that by the proper selection of the coefficients $c_k$
and angles $\theta_{k}$, it is possible to make all the off-diagonal
and diagonal elements of the new Toeplitz matrix $\sum_{k=1}^{2M+1}
c_k \mathbf D (\theta_{k})$ equal to zero except for a desired one.

First, note that the new Toeplitz matrix $\sum_{k=1}^{2M+1} c_k
\mathbf D (\theta_{k})$ can be expressed as $\mathbf T(\sum_{k=1}
^{2M+1} c_k \mathbf z(\theta_k))$. Based on the latter observation,
we conclude that all the off-diagonal elements of the matrix
$\sum_{k=1}^{2M+1} c_k \mathbf D (\theta_{k})$ are equal to zero
except for $j$th off-diagonal element if and only if all the
elements of the vector $\sum_{k=1}^{2M+1} c_k \mathbf z(\theta_k)$ 
are equal to zero except for the $j$th one. Therefore, let
us consider the following set of linear equations
\begin{equation}
\bigg [\mathbf z(\theta_1) \ \mathbf z(\theta_2) \  \cdots \
\mathbf z(\theta_{2M+1}) \bigg]  \mathbf c = \mathbf e_j .
\label{CoEq}
\end{equation}
where $\mathbf c \triangleq [c_1,\cdots,c_{2M+1}]$ and
$\mathbf e_j$ is the unit vector whose $j$th element is equal
to one. Since the newly defined matrix $\mathbf Z \triangleq
[\mathbf z(\theta_1) \ \mathbf z(\theta_2) \  \cdots \ \mathbf
z(\theta_{2M+1}) ]$ has the Vandermonde structure, it is
invertible under the condition that the angles $\theta_{k}, \
k=1,\cdots,2M+1$ are chosen to be distinct. 

In what follows, it is assumed that
\begin{equation}
\theta_{k}= (k-1)\frac{\pi}{2M+1}, \ k=1,\cdots, 2M+1
\end{equation}
and, thus, $\mathbf Z$ is invertible. Therefore, by selecting
the coefficient vector $\mathbf c$ in \eqref{CoEq} as $\mathbf 
c = \mathbf Z^{-1} \mathbf e_j, j=1,\cdots,2M+1$, the vector 
$\sum_{k=1}^K c_k \mathbf z(\theta_k)$ can be set equal to 
$\mathbf e_j$. It implies that in terms of the aforementioned 
selection of the coefficient vector $\mathbf c$ and angles 
$\theta_{k}, \ k=1,\cdots,2M+1$, it is possible to make all 
the off-diagonal and diagonal elements of the new Toeplitz 
matrix $\sum_{k=1}^K c_k \mathbf D (\theta_{k})$ equal to zero 
except for the $j$th one. Using \eqref{LinComb} and selecting 
$\mathbf c = \mathbf Z^{-1} \mathbf e_j, j=1, \cdots, 2M+1$, 
it can be resulted that $\sum_{k=1}^{2M+1} c_k \mathbf D 
(\theta_{k}) = \mathbf T (\mathbf e_j)$ which, in turn, 
implies that
\begin{equation}
\mathbf w^H \mathbf T (\mathbf e_j) \mathbf w = \mathbf 
v^H \mathbf T (\mathbf e_j)\mathbf v,\quad  j=1, \cdots, 2M+1. 
\label{Aux2}
\end{equation}
The set of equations in \eqref{Aux2} can equivalently be 
expressed as follows
\begin {eqnarray}
& & |w_1|^2+ |w_2|^2 + \cdots + |w_M|^2 = |v_1|^2+ |v_2|^2 + 
\cdots + |v_M|^2 \label{EQQ1} \\
& & w_1 w_2^* +  w_2 w_3^* + \cdots + w_{M-1} w_M^* = v_1 v_2^* 
+  v_2 v_3^* + \cdots + v_{M-1} v_M^* \\
& & w_1 w_3^* +  w_2 w_4^* + \cdots + w_{M-2} w_M^* = v_1 v_3^* 
+  v_2v_4^* + \cdots + v_{M-2}v_M^* \\
& & \vdots \quad \qquad \qquad \qquad \vdots \qquad \qquad \qquad 
\qquad \vdots \nonumber \\
& & w_1 w_{M-1}^* +  w_2 w_{M}^* = v_1 v_{M-1}^* +  v_2 v_{M}^* \\
& & w_1w_{M}^* = v_1v_{M}^* . \label{EQQ2}
\end {eqnarray}

Since any arbitrary Toeplitz matrix can be constructed by 
linearly combining the matrices $\mathbf T (\mathbf e_j), 
j=1,\cdots,2M+1$, equivalently, it can be concluded that by 
linearly combining the equations \eqref{EQQ1}--\eqref{EQQ2}, 
the equation \eqref{eq1} can be resulted. Thus, the 
beamforming vectors $\mathbf w$ and $\mathbf v$ have the 
same transmit beampattern if and only if the equations 
\eqref{EQQ1}--\eqref{EQQ2} are satisfied.

For a mother beamforming vector $\mathbf w$, we can construct
the set of all possible beamforming vectors that have the same
beampattern as $\mathbf w$ through solving the equations
\eqref{EQQ1}--\eqref{EQQ2}. For this goal, let us define the
following two functions of a single variable $x$
\begin{eqnarray}
f(x) \!\!\!\!\!&\triangleq&\!\!\!\!\! \overbrace{(w_1 + w_2 x + 
w_3 x^2 + \cdots + w_M x^{M-1})}^\text{First Multiplicative Term} 
\label{Aux4} \overbrace{(w_1^* + w_2^* x^{-1} + w_3^* x^{-2} 
+ \cdots + w_M^* x^{-M+1})}^\text{Second Multiplicative Term} \\
g(x) \!\!\!\!\!&\triangleq&\!\!\!\!\! (v_1 + v_2 x + v_3 x^2 + \cdots
+ v_M x^{M-1}) (v_1^* + v_2^* x^{-1} + v_3^* x^{-2} +
\cdots + v_M^* x^{-M+1}). 
\end{eqnarray}

By expanding the multiplicative terms in the definitions of $f(x)$ and
$g(x)$, it can be verified that the equations
\eqref{EQQ1}--\eqref{EQQ2} hold true if and only if $f(x)$ is equal to
$g(x)$. Let $x_0$ be a non-zero root of the first multiplicative term in
the definition of $f(x)$, i.e., $w_1 + w_2 x + w_3 x^2 + \cdots + w_M
x^{M-1}$. Then it is simple to verify that $1/x_o^*$ is also a root of
the second multiplicative term $w_1^* + w_2^* x^{-1} + w_3^* x^{-2} +
\cdots + w_M^* x^{-M+1}$ of $f(x)$. One implication of this observation 
is that the inverse conjugate of every root of $f(x)$ is also a root 
of $f(x)$ and, therefore, the roots of $f(x)$ can be denoted as $x_i$ 
and $1/x^*_i, \, i=1,\cdots, M-1$ and $f(x)$ can be decomposed as
\begin{eqnarray}
f(x) \!\!\!\!\!&=&\!\!\!\!\! |w_M|^2 \left( \frac{w_1}{w_M} + 
\frac{w_2}{w_M} x + \frac{w_3}{w_M} x^2 + \cdots +  x^{M-1} \right) 
\left( \frac{w_1^*}{w_M^*} + \frac{w_2^*}{w_M^*} x^{-1} + 
\frac{w_3^*}{w_M^*} + \cdots + x^{-M+1} 
\right) \nonumber \\
\!\!\!\!\!&=&\!\!\!\!\! |w_M|^2 \prod_{i=1}^{M-1}(x-x_i) \times
\prod_{i=1}^{M-1}(x^{-1}-x_i^*). \label{Aux3}
\end{eqnarray}
Furthermore, it is easy to verify that the product $(x-x_i) (x^{-1} -
x_i^*)$ can be equivalently expressed as
\begin{equation}
(x-x_i) (x^{-1}-x_i^*) = |x_i|^2 \left( x - \frac{1}{x_i^*} \right)
\left( x^{-1}-\frac{1}{x_i} \right) .
\end{equation}

Note that the product terms $\prod_{i=1}^{M-1}(x-a_i)$
and $\prod_{i=1}^{M-1}(x^{-1}-a_i^*)$ that appear in \eqref{Aux3}
preserve the structure of the first and second multiplicative 
terms in the definition of $f(x)$ ( see \eqref{Aux4}) for any 
arbitrary $a_i, i=1 , \cdots , M-1 $. Based on these observations, 
$f(x)$ can be decomposed as the multiplication of two terms in the 
form of $v_1 + v_2 x + v_3 x^2 + \cdots + v_M x^{M-1}$ and $v_1^* 
+ v_2^* x^{-1} + v_3^* x^{-2} + \cdots + v_M^* x^{-M+1}$ in 
$2^{M-1}$ different ways depending on whether $x_i$ (or $1/x^*_i,$) 
$i=1,\cdots, M-1$ is the root of the first polynomial. The 
corresponding coefficients of the first multiplicative term in each
decomposition correspond to a new beamforming vector dented as
$\mathbf v$ which has the same exact beampattern as $\mathbf w$.
This completes the proof. $\hfill\blacksquare$

\section{Application to Transmit Beamspace Design in MIMO Radar}
\begin{figure}[t]
\centerline{\includegraphics[width=10.0cm]{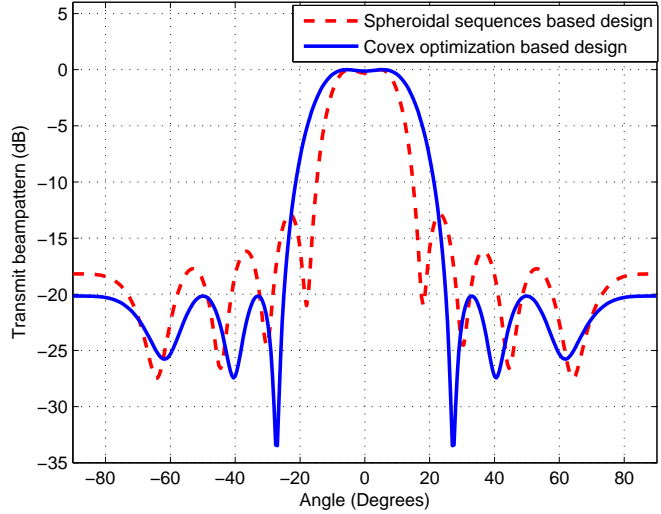}}
\caption{Transmit beampattern using spheroidal sequences and
convex optimization based designs.} \label{fi:Fig1}
\end{figure}

Consider a MIMO radar with transmit ULA of $M=10$ antennas spaced
half a wavelength apart. The total transmit power is normalized to
$P_t=M$. Two mother transmit beamforming weight vectors are
designed to focus the transmit energy within the sector $\boldsymbol
\Theta=[-10^{\circ},\ 10^{\circ}]$. The first mother beamforming weight
vector is designed using spheroidal sequences \cite{Foster},
\cite{Hassanien4}. Specifically, it is computed as ${\bf w}_{\rm SPH}
= \sqrt{P_t/2}({\bf u}_1 + {\bf u}_2)$, where ${\bf u}_1$ and ${\bf u}_2$
are the two principle eigenvectors of the matrix ${\bf A} =
\int_{\boldsymbol\Theta}{\bf a}(\theta){\bf a}^H(\theta) d \theta$. The
second mother beamforming weight vector is designed using convex
optimization to control the sidelobe levels. In particular, it is obtained
by solving the following convex optimization problem \cite{Hassanien4}
\begin{eqnarray}\label{eq:ConvexOpt}
\!\!\!\!\!\!&\!\!\!\!\!\!&\!\!\!\min_{\bf w} \max_{i} \|{\bf w}^H{\bf a}
({\theta}_i) - {e^{-j\phi_i}} \| ,\quad {\theta}_i \in {\boldsymbol\Theta}, 
\ i=1,\ldots,I \\
\!\!\!\!\!\!&\!\!\!\!\!\!&\!\!\!{\rm subject\ to}\ \|{\bf w}^H{\bf a}
({\theta}_k)\| \leq \delta , \quad {\theta}_k \in \bar{\boldsymbol\Theta},
\  k=1,\ldots,K
\label{eq:SidelobeLevels}
\end{eqnarray}
where $\bar{\boldsymbol\Theta}$ combines a continuum of all
out-of-sector directions, i.e., directions lying outside the
sector-of-interest $\boldsymbol\Theta$; $\phi_i, i=1,\ldots,I$ is the
desired transmit phase profile of user choice; and $\delta>0$ is the
parameter of the user choice that characterizes the worst acceptable
level of transmit power radiation in the out-of-sector region
$\bar{\boldsymbol\Theta}$. The phase $\phi_i=2\pi\sin(\theta_i)$
and the parameter $\delta=0.1$ are chosen, i.e, the sidelobe levels
are kept below $20\log{\delta}=-20$~dB. The mother beamforming
weight vector obtained by solving the problem
\eqref{eq:ConvexOpt}--\eqref{eq:SidelobeLevels} is referred to
hereafter as ${\bf w}_{\rm CVX}$. The transmit beampatterns
associated with mother beamforming weight vectors
${\bf w}_{\rm SPH}$ and ${\bf w}_{\rm CVX}$ are shown as the dotted
and solid curves in Fig.~\ref{fi:Fig1}, respectively.

The mother weight vector ${\bf w}_{\rm SPH}$ is used to generate a
population of $2^{M-1}=512$ weight vectors (including ${\bf w}_{\rm
SPH}$) of  dimension $M\times 1$. To implement a MIMO radar system
with four orthogonal transmit waveforms, the four beamforming weight
vectors among the population that achieve the best transmit power
distribution across the transmit array elements are chosen. The mother
weight vector ${\bf w}_{\rm SPH}$ and the four chosen weight vectors
denoted as ${\bf w}^{(j)}_{\rm SPH},\ j=1,\ldots,4$ are given in 
Table~1. Since the sector $\boldsymbol\Theta$ is symmetric around 0,
the mother beamforming vectors are real-valied. The four chosen 
vectors are scaled such that $\sum_j\|{\bf w}^{(j)}_{\rm SPH}\|^2
= P_t$. 

\begin{figure}[t]
\centerline{\includegraphics[width=10.0cm]{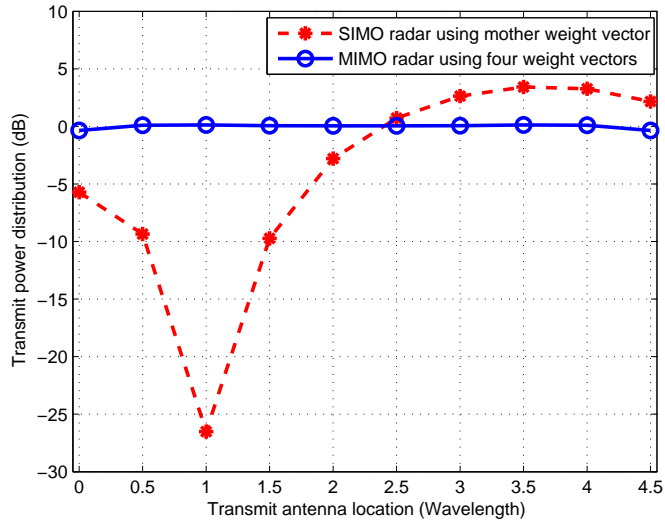}}
\caption{Transmit power distribution across the transmit array elements
for spheroidal sequences based transmit beamforming design.}
\label{fi:Fig2}
\end{figure}

It is worth noting that each of the four chosen vectors has the same 
beampattern as the mother beamforming vector except for the magnitude 
scaling factor of $1/4$. Note that the beampattern magnitude in the 
mainlobe as well as in the sidelobe regions is scaled by the same
scaling factor, i.e., the relative attenuation of the sidelobes with 
respect to the mainlobe remains unchanged. The transmit power 
distribution across the transmit array elements for the mother 
transmit beamforming weight vector operated in the SIMO radar mode and 
the four chosen weight vectors operated in the MIMO radar mode are shown 
in Fig.~\ref{fi:Fig2}. It can be seen from the figure that the

\ \\

{Table~1: Spheroidal sequences based mother transmit beamforming
vector and a subset of four vectors chosen from the $2^{M-1}$ 
population.\\ \ }
\begin{center}{
\small
\ \\
\begin{tabular}{|c|c|c|c|c|}
  \hline
  ${\bf w}_{\rm SPH}$ & ${\bf w}^{(1)}_{\rm SPH}$&${\bf w}^{(2)}_{\rm SPH}$
	&${\bf w}^{(3)}_{\rm SPH}$&${\bf w}^{(4)}_{\rm SPH}$ \\
    \hline
  \!\!\!0.5178  \!\!\!&\!\!\! 0.2589  \!\!\!&\!\!\! 0.2740 \!-\! 0.0000i  \!\!\!&\!\!\! 0.6061 \!-\! 0.0000i  \!\!\!&\!\!\! 0.6414\!\!\!\\
  \!\!\!0.3408  \!\!\!&\!\!\! 0.1704  \!\!\!&\!\!\! 0.1980 \!+\! 0.0255i  \!\!\!&\!\!\! 0.6490 \!-\! 0.0563i  \!\!\!&\!\!\! 0.7281\!\!\!\\
  \!\!\!0.0472  \!\!\!&\!\!\! 0.0236  \!\!\!&\!\!\! 0.0251 \!+\! 0.0475i  \!\!\!&\!\!\! 0.6791 \!-\! 0.1247i  \!\!\!&\!\!\! 0.7415\!\!\!\\
  \!\!\!-0.3263 \!\!\!&\!\!\! -0.1632 \!\!\!&\!\!\! -0.2122 \!+\! 0.0311i \!\!\!&\!\!\! 0.6797 \!-\! 0.1491i  \!\!\!&\!\!\! 0.6770\!\!\!\\
  \!\!\!-0.7253 \!\!\!&\!\!\! -0.3627 \!\!\!&\!\!\! -0.4458 \!-\! 0.0317i \!\!\!&\!\!\! 0.6098 \!-\! 0.1087i  \!\!\!&\!\!\! 0.5437\!\!\!\\
  \!\!\!-1.0873 \!\!\!&\!\!\! -0.5437 \!\!\!&\!\!\! -0.6098 \!-\! 0.1087i \!\!\!&\!\!\! 0.4458 \!-\! 0.0317i  \!\!\!&\!\!\! 0.3627\!\!\!\\
  \!\!\!-1.3540 \!\!\!&\!\!\! -0.6770 \!\!\!&\!\!\! -0.6797 \!-\! 0.1491i \!\!\!&\!\!\! 0.2122 \!+\! 0.0311i  \!\!\!&\!\!\! 0.1632\!\!\!\\
  \!\!\!-1.4830 \!\!\!&\!\!\! -0.7415 \!\!\!&\!\!\! -0.6791 \!-\! 0.1247i \!\!\!&\!\!\! -0.0251 \!+\! 0.0475i \!\!\!&\!\!\! -0.0236\!\!\!\\
  \!\!\!-1.4562 \!\!\!&\!\!\! -0.7281 \!\!\!&\!\!\! -0.6490 \!-\! 0.0563i \!\!\!&\!\!\! -0.1980 \!+\! 0.0255i \!\!\!&\!\!\! -0.1704\!\!\!\\
  \!\!\!-1.2828 \!\!\!&\!\!\! -0.6414 \!\!\!&\!\!\! -0.6061 \!-\! 0.0000i \!\!\!&\!\!\! -0.2740 \!+\! 0.0000i \!\!\!&\!\!\! -0.2589\!\!\!\\
  \hline
\end{tabular}} 
\end{center}
\ \\

\!\!\! mother
beamforming weight vector has very poor transmit power distribution.
For example, the power radiated by the third transmit array element
is over 25~dB less than the average transmit power per transmit array
element. On the other hand, the four chosen beamforming vectors exhibit 
transmit power distribution that is almost uniform, which is desirable 
in practice.

\begin{figure}[t]
\centerline{\includegraphics[width=10.0cm]{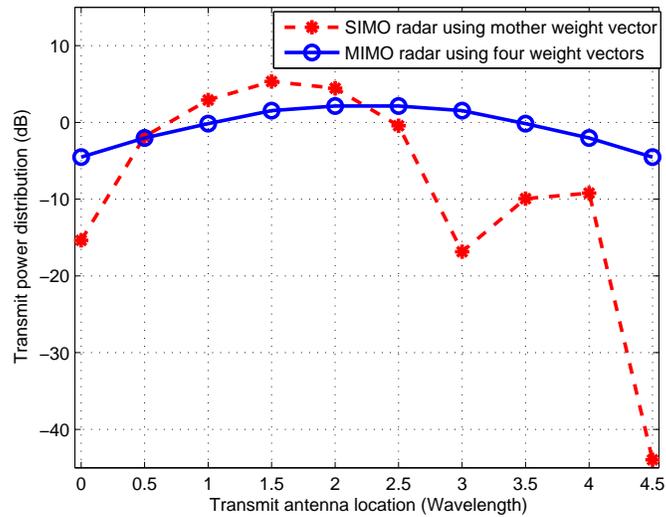}}
\caption{Transmit power distribution across the transmit array elements
for convex optimization based transmit beamforming design.}
\label{fi:Fig3}
\end{figure}

Similarly, the mother weight vector ${\bf w}_{\rm CVX}$ is used to
generate a population of $2^{M-1}=512$ beamforming vectors
(including ${\bf w}_{\rm CVX}$) which have the exact same
beampattern. The four beamforming weight vectors among the
population that achieve the best transmit power distribution across
the transmit array elements are chosen. The mother weight vector
${\bf w}_{\rm SPH}$ and the four chosen weight vectors denoted
as ${\bf w}^{(j)}_{\rm CVX},\ j=1,\ldots,4$ are given in Table~2. The
transmit power distributions across the transmit array elements for
the mother weight vector and the four chosen weight vectors are
shown in Fig.~\ref{fi:Fig3}. It can be seen from the figure that the
four chosen vectors yield much better transmit power distribution
as compared to the mother weight vector.
\ \\

\newpage
{Table~2: Convex optimization based mother transmit beamforming vector 
and a subset of four vectors chosen from the $2^{M-1}$ population.\\ \ }
\begin{center}{
\small
\begin{tabular}{|c|c|c|c|c|}
  \hline
  ${\bf w}_{\rm CVX}$ & ${\bf w}^{(1)}_{\rm CVX}$&${\bf w}^{(2)}_{\rm CVX}$
	&${\bf w}^{(3)}_{\rm CVX}$&${\bf w}^{(4)}_{\rm CVX}$ \\
    \hline
  -0.1702  \!\!\!&\!\!\! 0.0011  \!\!\!&\!\!\! 0.0005   \!\!\!&\!\!\! 0.5412   \!\!\!&\!\!\! 0.2438 \\
  0.8093   \!\!\!&\!\!\! -0.0661 \!\!\!&\!\!\! -0.0303  \!\!\!&\!\!\! 0.7829   \!\!\!&\!\!\! 0.0930 \\
  1.4003   \!\!\!&\!\!\! 0.2481  \!\!\!&\!\!\! 0.1432   \!\!\!&\!\!\! 0.9311   \!\!\!&\!\!\! -0.1178 \\
  1.8485   \!\!\!&\!\!\! 0.6505  \!\!\!&\!\!\! 0.1931   \!\!\!&\!\!\! 0.7435   \!\!\!&\!\!\! -0.6407 \\
  1.6711   \!\!\!&\!\!\! 0.8511  \!\!\!&\!\!\! 0.0333   \!\!\!&\!\!\! 0.3386   \!\!\!&\!\!\! -0.8910 \\
  0.9506   \!\!\!&\!\!\! 0.8910  \!\!\!&\!\!\! -0.3386  \!\!\!&\!\!\! -0.0333  \!\!\!&\!\!\! -0.8511 \\
  0.1437   \!\!\!&\!\!\! 0.6407  \!\!\!&\!\!\! -0.7435  \!\!\!&\!\!\! -0.1931  \!\!\!&\!\!\! -0.6505 \\
  -0.3186  \!\!\!&\!\!\! 0.1178  \!\!\!&\!\!\! -0.9311  \!\!\!&\!\!\! -0.1432  \!\!\!&\!\!\! -0.2481 \\
  -0.3459  \!\!\!&\!\!\! -0.0930 \!\!\!&\!\!\! -0.7829  \!\!\!&\!\!\! 0.0303   \!\!\!&\!\!\! 0.0661 \\
  0.0063   \!\!\!&\!\!\! -0.2438 \!\!\!&\!\!\! -0.5412  \!\!\!&\!\!\! -0.0005  \!\!\!&\!\!\! -0.0011 \\
  \hline
\end{tabular}
}
\end{center}

\newpage
\section{Conclusion}
The important question regarding the existence of other beamforming
vectors whose transmit beampatterns are the exact same as the transmit
beampattern of a given beamforming vector has been addressed. It has
been proven that at most $2^{M-1}-1$ other beamforming vectors with
the same beampattern can be generated from any given beamforming
vector. The method for constructing the set of all possible beamforming
vectors from a given mother beamforming vector has been also
developed. Moreover, it has been shown how this result can be utilized
in the actively developing field of transmit beamspace design for MIMO
radar systems, where it is desirable to have different transmit waveforms
to be radiated with the same transmit beampattern.


\begin{thebibliography}{10}
\bibitem{Anderson} S.~Anderson, ``On optimal dimension reduction for
sensor array signal processing,'' {\it Signal Processing}, vol.~30,
pp.~245--256, Jan.~1993.

\bibitem{Foster} P.~Forster and G.~Vezzosi, ``Application of spheroidal
sequences to array processing,'' in {\it Proc. IEEE Int. Conf. Acoustics,
Speech, Signal Processing}, Dallas, TX, May~1987, pp.~2268--2271.

\bibitem{Vorobyov} S.~A.~Vorobyov, ``Adaptive and robust beamforming,'' in
{\it Academic Press Library in Signal Processing, Vol. 3, Array and
Statistical Signal Processing}, Eds. R. Chellappa and S. Theodoridis,
Academic Press, 2014, pp.~503--552.

\bibitem{VanTrees} H.~L.~Van~Trees, {\it Optimum Array Processing}.
NewYork: Wiley, 2002.

\bibitem{Gershman} A.~B.~Gershman, ``Direction finding using beamspace
root estimator banks,'' \emph{IEEE Trans. Signal Processing}, vol.~46, no.~11,
pp.~3131--3135, Nov.~1998.

\bibitem{Zoltowski} M.~D.~Zoltowski, G.~M.~Kautz, and S.~D.~Silverstein,
``Beamspace root-MUSIC,'' \emph{IEEE Trans. Signal Processing}, vol.~41,
no.~1, pp.~344--364, Jan.~1993.

\bibitem{Farina} A.~Farina, {\it Antenna Based Signal Processing
Techniques for Radar Systems.} Norwood, MA: Artech House, 1992.

\bibitem{Fuhrman} D.~R.~Fuhrmann and G.~San~Antonio, ``Transmit
beamforming for MIMO radar systems using partial signal correlation,''
in {\it Proc. Asilomar Conf. Signals, Syst., Comput., Asilomar}, CA,
USA, Nov.~2004, pp.~295--299.

\bibitem{Stioca} P.~Stoica, J.~Li, and Y.~Xie, ``On probing signal
design for MIMO radar,'' IEEE Trans. Signal Processing, vol.~55,
no.~8, pp.~4151--4161, Aug.~2007.

\bibitem{Fuhrman2} D.~R.~Fuhrmann and G.~San~Antonio, ``Transmit
beamforming for MIMO radar systems using signal cross-correlation,''
{\it IEEE Trans. Aerospace and Electronic Systems}, vol.~44,
pp.~171--186, Jan.~2008.

\bibitem{Chen} C.-Y.~Chen and P.~Vaidyanathan, ``MIMO radar space-time
adaptive processing using prolate spheroidal wave functions,'' {\it IEEE
Trans. Signal Processing}, vol.~56, no.~2, pp.~623--635, Feb.~2008.

\bibitem{Browning} J.~P.~Browning, D.~R.~Fuhrmann, and M.~Rangaswamy,
``A hybrid MIMO phased-array concept for arbitrary spatial beampattern
synthesis,'' in Proc. {\it IEEE Digital Signal Processing Signal
Processing Education Workshop (DSP/SPE)}, Marco Island, FL, Jan.~2009,
pp.~446--450.

\bibitem{Fuhrman3} D.~R.~Fuhrmann, P.~Browning, and M.~Rangaswamy,
``Constant-modulus partially correlated signal design for uniform
linear and rectangular MIMO radar arrays,'' in {\it Proc. 4th Int.
Conf. Waveform Diversity Design (WDD)}, Orlando, FL, Feb.~2009,
pp.~197--201.

\bibitem{Hassanien3} A.~Hassanien and S.~A.~Vorobyov, ``Transmit/receive
beamforming for MIMO radar with colocated antennas,'' in {\it Proc. IEEE
Int. Conf. Acoustic, Speech, Signal Processing (ICASSP)}, Taipei,
Taiwan, Apr.~2009, pp.~2089-2092.

\bibitem{Hassanien1} A.~Hassanien and S.~A.~Vorobyov, ``Direction finding
for MIMO radar with colocated antennas using transmit beamspace
preprocessing,'' in {\it Proc. 3rd IEEE Inter. Workshop Computational
Advances in Multi-Sensor Adaptive Processing}, Aruba, Dutch Antilles,
Dec.~2009, pp.~181--184.

\bibitem{Aittomaki} T.~Aittomaki and V.~Koivunen, ``Beampattern
optimization by minimization of quartic polynomial,'' in {\it Proc.
15 IEEE/SP Statist. Signal Processing Workshop}, Cardiff, U.K.,
Sep.~2009, pp.~437--440.

\bibitem{Fuhrman4} D.~R.~Fuhrmann, P.~Browning, and M.~Rangaswamy,
``Signaling strategies for the hybrid MIMO phased-array radar,'' {\it
IEEE J. Sel. Topics Signal Processing}, vol.~4, no.~1, pp.~66--78,
Feb.~2010.

\bibitem{Hassanien5} A.~Hassanien and S.~A.~Vorobyov, ``Phased-MIMO
radar: A tradeoff between phased-array and MIMO radars,'' {\it IEEE
Trans. Signal Processing}, vol.~58, no.~6, pp.~3137--3151, June~2010.

\bibitem{Hassanien4} A.~Hassanien and S.~A.~Vorobyov, "Transmit energy
focusing for DOA estimation in MIMO radar with colocated antennas,"
{\it IEEE Trans. Signal Processing}, vol.~59, no.~6, pp.~2669--2682,
June~2011.

\bibitem{Hassanien2} A.~Hassanien and S.~A.~Vorobyov, ``Subspace-based
direction finding using transmit energy focusing in MIMO radar with
colocated antennas,'' in {\it Proc. IEEE Int. Conf. Acoust., Speech,
Signal Processing}, Prague, Czech Republic, May~2011, pp.~2788--2791.

\bibitem{Khabbazi} A.~Khabbazibasmenj, S.~A.~Vorobyov, and A.~Hassanien,
``Transmit beamspace design for direction finding in colocated MIMO radar
with arbitrary receive array,'' in {\it Proc. 36th IEEE Inter. Conf.
Acoustics, Speech, and Signal Processing}, Prague, Czech Republic,
May~2011, pp.~2784--2787.

\bibitem{Tse} D.~Tse and P.~Viswanath, {\it Fundamentals of Wireless
Communication}. Cambridge University Press, 2005.

\bibitem{Goldsmith} A.~Goldsmith, {\it Wireless Communications}.
Cambridge University Press, 2005.

\bibitem{Molish} A.~F.~Molisch, {\it Wireless Communications}.
John Wiley and Sons, 2010.

\bibitem{Jieping} Y.~Jieping, L.~Qi, X.~Hui, H.~Park, R.~Janardan,
and V.~Kumar, ``IDR/QR: An incremental dimension reduction algorithm
via QR decomposition,'' {\it IEEE Trans. Knowl. Data Eng.}, vol.~19,
no.~9, pp.~1208--1222, Sep.~2005.

\bibitem{Haiping} L.~Haiping, K.~Plataniotis, and A.~Venetsanopoulos,
``MPCA: Multilinear principal component analysis of tensor objects,''
{\it IEEE Trans. Neural Networks}, vol.~19, no.~1, pp.~18--39,
Jan.~2008.

\bibitem{DimRed} A.~Hassanien and S.~A.~Vorobyov, ``A robust adaptive 
dimension reduction technique with application to array processing,'' 
{\it IEEE Signal Processing Letters}, vol.~16, no.~1, pp.~22-25, 
Jan.~2009.

\bibitem{Rivera} A.~Rodriguez-Rivera, B.~Baryshnikov, B.~Van~Veen,
and R.~Wakai, ``MEG and EEG source localization in beamspace,'' {\it
IEEE Trans. Biomedical Engineering}, vol.~53, no.~3, pp.~430--341,
Mar.~2006.

\bibitem{Roy} R.~Roy and T.~Kailath, ``ESPRIT estimation of signal
parameters via rotational invariance techniques,'' {\it IEEE Trans.
Acoustic, Speech, Signal Processing}, vol.~37, no.~7, pp.~984--995,
Jul.~1989.

\bibitem{KHVM} A.~Khabbazibasmenj, A.~Hassanien, S.~A.~Vorobyov,
and M.~W.~Morency, ``Efficient transmit beamspace design for
search-free based DOA estimation in MIMO radar,'' {\it IEEE Trans.
Signal Processing}, vol.~62, to be published in 2014.
(http://arxiv.org/abs/1305.4979)

\bibitem{WaveletBook} M.~Vetterli and J.~Kovacevic. {\it Wavelets
and Subband Coding}. Vol. 87. Englewood Cliffs, New Jersey: Prentice
Hall PTR, 1995.
\end{thebibliography}
\end{document}